# Individual versus collective cognition in social insects

Ofer Feinerman*, Amos Korman¥

* Department of Physics of Complex Systems, Weizmann Institute of Science, 7610001, Rehovot, Israel. Email: ofer.feinerman@weizmann.ac.il

¥ Institut de Recherche en Informatique Fondamentale (IRIF), CNRS and University Paris Diderot, 75013, Paris, France. Email: amos.korman@liafa.univ-paris-diderot.fr

## Abstract

*The concerted responses of eusocial insects to environmental stimuli are often referred to as collective cognition on the level of the colony.To achieve collective cognitiona group can draw on two different sources: individual cognitionand the connectivity between individuals.Computation in neural-networks, for example,is attributedmore tosophisticated communication schemes than to the complexity of individual neurons. The case of social insects, however, can be expected to differ. This is since individual insects are cognitively capable units that are often able to process information that is directly relevant at the level of the colony.Furthermore, involved communication patterns seem difficult to implement in a group of insects since these lack clear network structure.This review discusses links between the cognition of an individual insect and that of the colony. We provide examples for collective cognition whose sources span the full spectrum between amplification of individual insect cognition and emergent group-level processes.*

## Introduction

The individuals that make up a social insect colony are so tightly knit that they are often regarded as a single super-organism(Wilson and Hölldobler, 2009). This point of view seems to go far beyond a simple metaphor(Gillooly et al., 2010)and encompasses aspects of the colony that are analogous to cell differentiation(Emerson, 1939), metabolic rates(Hou et al., 2010; Waters et al., 2010), nutrient regulation(Behmer, 2009),thermoregulation(Jones, 2004; Starks et al., 2000), gas exchange(King et al., 2015), and more.

It is tempting to push this analogy, one step further and attribute the superorganism with collective cognition(Couzin, 2009; Franks, 1989; Seeley, 1996). In this respect, it is possible to envision two extreme cases in which groups of insects may have evolved to exhibit cognition on the scale of the entire colony. The first is reliance on the cognition of the individuals that make up the group. Indeed, the cognitive abilities of a single ant or bee within the large colony are far from being simple (Dornhaus and Franks, 2008). The group can benefit from these capabilities, for example, by sharing and refining the knowledge of informed individuals. The second extreme case is collective cognition derived from the interaction between members. Manmade systems teach us that complex computation can be achieved by the wiring together of very simple components such as logical gates (Lindgren and Nordahl, 1990). Similarly, social insect colonies often display dense interaction networks (Wilson and Hölldobler, 1988)and collective behaviors that appear to exceed the capacity of the individuals of which they are comprised (Sumpter, 2006).

It is therefore of interest to trace the collective actions of the social insect colony to their sources, be they the cognition of individuals or the communication network that connect groups of such individuals. Mapping out the relations between these two organizational scales is required if one is to understand and quantify collective cognition as well as learn about its evolutionary origins.We hypothesize that individual-based collective behaviors will be prevalent in cases where abilities, similar to those exhibited by solitary insects, suffice in order to sense, grasp, and process knowledge that is relevant on the scale of the colony. Deviations from this will tend to lead to group solutions that involve an increased emergent component.

## Outline

The outline of this review is as follows: First, we discuss the cognitive abilities of the individuals that make up the social insect colony andsome current knowledge of communication networks in social insects. As mentioned above, these two components provide the basis on which the colony could build its collective capabilities. Next, we present a list of examples of collective cognition. These examples are ordered by the degree to which collective behaviors rely on each of the two components, from individual-based to connectivity-based. The examples are split into three categories: Individual-based collective behaviors, collective behaviors that combine different individual perspectives, and, finally, collective behaviors that display higher levels of emergence. Each of these categories is divided into subcategories that further refine this division. Taken together, these examples span a broad spectrum of relationships between individual and collective cognition. In the final section we discuss the possible factors that may determine the degree of emergence in a particular collective behavior.

## Individual cognition

A good starting point for discussing the origins of cognition in social insect colonies is the cognitive abilities of solitary insects. Insect brains have evolved hundreds of millions of years(Farris and Schulmeister, 2010; Ma et al., 2012)prior to the appearance of eusociality(Moreau, 2006). Despite the fact that their brains are relatively small(Chittka and Niven, 2009),solitary insects exhibit high cognitive skills that include large behavioral repertoires(Evans, 1966), complex forms of learning(Alloway, 1972; Blackiston et al., 2011), and include navigational skills that often exceed those of humans(Brower, 1996). These abilities aid the solitary insect, among other things, inforaging (O'Neill, 2001), finding or constructing shelters(Raw, 1972), confronting predators(Schmidt, 1990), and identifying appropriate mating partners(Dickson, 2008).

The next step in this discussion is the transitions to eusociality which happened between100-150 million years ago(Brady et al., 2006; Engel et al., 2009). Eusociality is characterized by reproductive division of labor that drastically lowers the level of conflict between group members as they strive towards common goals (Crespi and Yanega, 1995). It is first important to state the evident fact that, contrary to cells in a tissue or neurons in the brain, insects within the colony superorganism maintain their individuality. They are able of autonomous motion and decision making.Further, the brains of individuals within a colony bear high resemblance to those of solitary insects(Strausfeld, 1976). One may therefore ask how the cognitive capabilities a social insect compare to those of a solitary insect. To date, it is not clear if once grouped into large groups evolution may work to increase or decrease the cognitive complexity of individuals. On the one hand, it is known, mainly from vertebrate groups that the communication requirements of group living may work to increase brain complexity(Shultz and Dunbar, 2010). On the other hand, it has been suggested that, relying on collective processes may ease the energetically expensive(Aiello and Wheeler, 1995) maintenanceof brain tissue(Anderson and McShea, 2001; Feinerman and Traniello, 2015).

Whatever the exact comparison between the brains of a social and a solitary insect, it is clear that thesocial insect is a cognitively capable individual. Individuals within the colony possess the capacity for large behavioral repertoires(Chittka and Niven, 2009), for weighing a large number of factors to reach individual decisions(Franks et al., 2003), andfor navigating over large distances(Gathmann and Tscharntke, 2002; Wehner, 2003). Importantly, these individual capabilities are relevant on the scale of the entire colony.

One aspectthat clearly differentiates the social insect from its solitary counterpart is the capacity for communication. For example, eusocial insects display a huge diversification of cuticular pheromones(van Wilgenburg et al., 2011) used to convey multiple signals that are unique to colony life(Howard and Blomquist, 2005). Another famous example is the honeybee waggle dance(Von Frisc, 1950). While solitary insects may have the motivation to conceal a newly found item for personal consumption(Byrne et al., 2003), the bees have evolved an elaborate communication scheme which allows them to share this location.These and other interaction skills form the foundation of the insect society. In the next section we discuss some of the properties of the communication networks via which social insects coordinate their activities.

**Interaction Networks**

Group living animals combine personal and social information when deciding upon their next action(Rieucau and Giraldeau, 2011). In eusocial insects – the social component of information collection is especially important(Wilson and Hölldobler, 1988). Correspondingly, the modalities of communication and richness of cues and signals is greatly enhanced. Social insects use a variety of olfactory(Martin and Drijfhout, 2009; Morgan, 2009), tactile(Razin et al., 2013), visual, and vibrational(Delattre et al., 2015; Roces et al., 1993)messages as well as multi-model combinations of these (Ramsden et al., 2009) in their communication. Broadly speaking, these can be divided into several groups: Some messages require direct contact between individuals and can thus be considered as local in both space and time. Other signals are local in time but not space and are typically employed as alarm signals (e.g. highly volatile pheromones(Blum, 1969)). Yet another group are signals that are local in space but not in time. This group includes stigmergic, indirect communication between insects in which one individual modifies the environment and a second individual arriving at the same location at some later time reacts to its modified surroundings(Theraulaz and Bonabeau, 1999). Mass recruitment pheromone trails (Jaffe and Howse, 1979)and nest construction without a blueprint (Franks and Deneubourg, 1997)are two impressive examples of stigmergy. Note that, for the case of pheromonal communication, the time scales that characterize pheromonal communication are evolvable as they depend on chemical evaporation times that,indeed, vary between species and tasks(Morgan, 2009; Witte et al., 2007)(Holldobler and Wilson, 1990).

Quantifying the communication patterns requires descriptive frameworksfor the different interaction types as described above. Contact dependent interactions can be described as time-ordered (Blonder and Dornhaus, 2011)communication networks(Fewell, 2003; Moreau et al., 2011). It has been shown that, in a laboratory setting, the high mobility of social insectsdictates that, after a sufficient time window, interactions occur between practically all possible pairs and the network become highly connected (Mersch et al., 2013). Stigmergic communication has been described by using the language of statistical mechanics (Richardson et al., 2011), or by employing cellular-automata tools typically used to describe self-organization processes (Khuong et al., 2016). These interactions are, inherently, one-to-many signaling and have been shown to extend the connectivity induced by contact dependent communication (Richardson and Gorochowski, 2015)..Adding long range communication such as that involving alarm pheromones, we obtain a picture of a system in which, at least to first order and over long enough time-scales, interactions can be described as well mixed. In other words, over time an insect receives signals from any other insect in the colony. Not only are interactions mixed they are also, to a large extent anonymous. With a few exceptions (Mallon and Franks, 2000; Tibbetts, 2002), it is reasonable to assume that individuals do not recognize which of the hundreds to hundreds of thousands of other individuals they are currently interacting with.

While the previous discussion seems to suggest that interactions are completely ergodic, it is important to stress that the social insect colony is, by no means, devoid of structure. For example, ant nests and bee hives are often concentrically arrangedsuch that young insects reside in the deep center while older individuals occupy progressively occupy areas that are closer to the boundaries or entrance (Beshers and Fewell, 2001). Even when structure is initially lacking, self-organization and amplification of noise can work to create spatio-temporal patterns over time (Richardson et al., 2011; Theraulaz et al., 2003). Moreover, ant (Tschinkel, 2004) and termite (Noirot and Darlington, 2000) nests exhibit complex structures of rooms and corridors and these further reflect on the spatial distribution of individuals within the nest. It has been shown that different individuals tend to occupy specific locations or chambers within the nest (Jandt and Dornhaus, 2009; Sendova-Franks and Franks, 1995) and that this reflects on the probability that they interact with other individuals in other parts of the nest (Mersch et al., 2013; Pinter-Wollman et al., 2011). Spatial locations therefore inducea network structure that is composed of relatively stable clusters. Hence, in a very broad sense, the communication patterns in a social insect colony can be viewed as residing between a well-mixed(on the more local scale) and a fixed (on the global, cluster, scale) network.

## Collective cognition

Having described some of the basic "cognitive toolbox" available to the colony, we go on to discuss its collective scale behavioral products. In what follows, we presenta non-comprehensive list of examples for collective behaviors in social insect colonies.

Theexamples where chosen while focusing on the different possible gaps between the knowledge, actions, and capabilities of individuals and those of the entire group. They are ordered by the extent of this gap and divided between individual based collective cognition (small gaps) and emergent collective cognition that builds on the interaction between insects (large gaps). This division is, by no means, strict.

## *Individual based collective cognition*

In a eusocial colony the genetic conflict between individuals in the group is minimal(Queller and Strassmann, 1998). This leads to an alignment of interests which implies that it is generally advantageous for informed individuals to share their knowledge with other group members. Utilizing this information is useful for the group as well. Since the number of informed individuals may be small one could expect that their actions be too weak to elicit any significanteffect or, alternatively, they be averaged out against opposing actions performed by other, less-informed, colony members. Instead of losing this useful information, it may be profitable for the group to amplify the actions of these focal individuals. Anysuch amplification should be regulated to prevent runaway behavior in case of mistakes. Such mistakes could arise from the informed individuals themselves: they may hold only partial information or be plain wrong, or from communication: noisy interactions may distort the original message.

Next, we discuss several examples ofamplification circuits that make the products of individual cognitionavailable, effective, anduseful at the level of the group.

**Unconditional amplification**
The simplest example is the alarm response. When an individual ant senses danger she not only directly reacts to it but further emits a volatile alarm pheromone(Blum, 1969). This pheromone spreads around the ant eliciting similar responses from her neighboring nest-mates. This positive feedback circuit quickly spreads the danger signal to affect a large number of individuals(Jeanson and Deneubourg, 2009). This not only increases the group's surveillance of its environment (the "many eyes principle") but also allows it to take collective actions towards, for example, protection of the nest. Similar behaviors are displayed by termiteswhere chemical communication is accompanied by vibrationalsignaling (Delattre et al., 2015).

These collective positive feedback circuits provide an informed individual that senses danger immediate and direct control over the actions of the group. In other words,the gap between individual and collective cognition is, practically, nonexistent.The group forsakes regulation and out-weights this crucial survival response over the possible price paid by false alarms.

**Conditioned amplification**
The mass recruitment foraging trail occurs as a single first ant locates a food source. This ant then uses her navigational skills to return to the nest while laying a pheromone trail that recruits others to the food such that foraging commences. While an emergent process may work to straighten the trail and make it shorter (see below) the trail still follows the qualitative solution as first discovered and then communicated by the initial recruiter(Holldobler, 1971).

Importantly, in a large number of species, ants strengthen the initial trail only on the way back from the target food source and only if they independently found it to be profitable(Beckers et al., 1992a; Mailleux et al., 2003; Wilson, 1962). This regulationcan be considered as "delayed" since it occurs only after an initial positive response to the initial ant. This provides a mechanism by which the group"double-checks"the target communicated by the initial ant before continuing to amplify her effect even further. Itallows the colony to reduce its response to ants that may have outdated information or are, for some reason, confused. Since pheromones are volatile and have a finite lifetime, their concentration along the trail depends on the rate at which they are enhanced. A trail which is not enhanced will eventually disappear. Thus, delayed regulation further supplies a mechanism for calibrating the level of activity on the recruitment trail (Simon and Hefetz, 1992), discontinuingit once the food source is exhausted(Wilson, 1962). It further allows the system to escape local minima by switching to foraging on more profitable food sources when such are identified(Beekman and Dussutour, 2007).

**Amplification with early regulation**

Desert ants typically forage alone and display only a rudimentary form of recruitment(Amor et al., 2010; Razin et al., 2013). The recruitment process occurs as ants that are informed about a food source outside the nest attempt to alert their nest-mates using imperfect communication. The interactions used are noisy in the sense that a recruitment interaction may be ignored or, conversely, a non-recruitment interaction may induce an ant to leave the nest. Therefore, amplification of the initial signal must be regulated so that recruitment occurs only when a food source is present and runaway behavior that results from mere interaction noise is avoided. It was shown that desert ants regulate recruitment early on in the process, at the entrance chamber of their nest(Razin et al., 2013).

This regulation is the result of two behavioral components. The firstis the fact that individuals "know that they know"(Greenwald et al., 2015). There is a clear difference in the way directly and non-directly informed insects react to interactions with others(Razin et al., 2013; Schultz et al., 2008; Stroeymeyt et al., 2011). For example, ants that have been to the food areasimply disregard interactions and maintain high walking speed to increase the number and effectiveness of their recruitment interactions(Razin et al., 2013). On the other hand, ants with second hand knowledge react to interactions in a more cautious manner. These ants can either upregulate or downregulate their propensity to be recruited depending on the state of the individual they interact with(Razin et al., 2013). These rules allow the ants to regulate collective behavior with minimaldependence on theirunreliablecommunication skills. Specifically, thisworks to decrease the chances that the actions of non-informed ants have global consequences on the state of the nest and leaves the stage, so to say, to the directly informed ants(Razin et al., 2013).

The second regulatory component is an early negative feedback. All else being equal, non-informed ants tend to lower their propensity to exit with passing time. This leads to a collective threshold that dissipates the effects of random or isolated interactions such that their effect quickly dies away and this protects the system from noise. A persistent informed ant with first-hand information can generate enough activity such that interaction rates increase(Gordon and Mehdiabadi, 1999) and the system moves over this recruitment threshold. This early feedback mechanism in which positive feedback occurs only if the system passes a set threshold is similar to the generation of spikes in neurons(Razin et al., 2013). Note that while delayed regulation (as described in the previous subsection) works to regulate the amplitude of the collective response and terminate it when the stimulus ends, early regulation works to prevent amplification in the first place.

## *Combining individual perspectives*

### Amplifying the optimal option

The social insect colony may do more than amplifying individual decisions – it can, in fact, poll individuals to reach consensus choice regarding the best solution among several alternative options. Examples for this come from house-hunting(Visscher, 2007) behaviors in ants(Franks et al., 2002) and bees(Seeley et al., 2006). When assessing the quality of a potential new nest site scout ants have been shown to incorporate an intricate, individually based,evaluation scheme which combines the different attributes of this location (e.g. its volume, the size of the door, and the level of light) in a non-trivial way(Franks et al., 2003).There is evidence that this assessment results in a single grade given by the ant to the new location(Robinson et al., 2011; Robinson et al., 2014). The group does not make its own assessments of nest quality (a hypothetical example for this could be moving the whole colony to occupy the alternative nests one at a time and using the resulting ant density (Gordon et al., 1993)to accurately measure the area of each) but, rather, uses a quorum sensing as a polling mechanism to compare the assessments of its individual scouts(Franks et al., 2002; Seeley et al., 2006). With high probability, this leads to the colony choosing the best among the alternatives with the accuracy of the decision growing with the size of the group(Sasaki et al., 2013).

House-hunting provides another fascinating example of how the action of the group may work to refine individual decisions: When comparing different nests with specific attributes individual ants are prone to violate theregularity principle of rational decision making(Sasaki and Pratt, 2011). This principle states that if option A is preferred over option B then this should not change upon introducing a third option, C, that is inferior to both. This fallacy is not specific to insects but, rather, affects many different animals including humans. However, when a whole colony is presented with the choice between these nests it will tend to make the rational choice(Edwards and Pratt, 2009; Sasaki and Pratt, 2011). This is because the polling often terminates before individual ants have had the chance to fail the regularity principle since this requires visiting multiple nests(Robinson et al., 2014).

We provided several examples (many more exist) of how the actions and decisions of capable individuals reflect at the level of the group. The group does not create new solutions but ratherworks to amplify, average, poll, and refine the actions of individual members. This is done using collective communication circuits that involve positive and negative feedbacks and certain non-linearities.

### Amplification in dynamic settings

In dynamic, fast-evolving scenarios, Information can quickly become obsolete and the individuals that carry useful information (Robson and Traniello, 2002) change over time (Gelblum et al., 2015). This entails two inherent problems: First, the relevant individual at a specific point in time has to be identified by the group. Second, when an individual that is better updated appears the group must revert to following it instead.

This scenario is realized during cooperative transport by longhorn crazy ants (Czaczkes and Ratnieks, 2013). When ants cooperatively transport a large food item they can often lose orientation and become unknowledgeable regarding the correct way to the nest. To correct their path, these ants rely on well-informed individuals that are in the vicinity of the load but unattached to it (Gelblum et al., 2015).

Instead of identifying the ant that currently holds valuable navigational information, here again, the group relies on the "know that you know" principle. In other words, an informed ant acts in a manner that is different from the other carriers: She attaches to the object and, without heeding to others, pulls it in the direction she knows to be correct. At the same time the carrying ants "acknowledge that they don't know" and apply a different behavioral rule – which is, in a sense, pull in the direction in which the load is currently moving. Together, these different rules as applied by informed and non-informed ants allow the group to optimally amplify the force of the informed leader (Gelblum et al., 2015).

Importantly, after a period of about 10 seconds, the newly attached leader loses her orientation. This former leader then adapts the behavioral rules of an ordinary carrier and may continue in this state for many minutes. The directionality of the carrying group is next corrected by the attachment of a new leader ant that happened to be informed at that particular time. The fast switching between leaders can be viewed as a mechanism that enables the group to escape being trapped at local minima in which the group displays coordinated motion but the direction is wrong.

**Collective bootstrapping of individual solutions**
Another form of colony level solutions that is based on the actions of a large number of individuals is trail shortening. Ants are famous for the ability to gradually decrease the length of their pheromone trail so that it finally draws a geodesic between the food source and the nest (Feynmann, 1985). This process can occur by the accumulated effect of ants that leave the trail (Deneubourg et al., 1983) and return to it a short distance away. Useful detours, *i.e.* those that "cut a corner" and slightly decrease the trail's length, are then amplified by the group while non-useful detours are abandoned (Deneubourg et al., 1983; Goss et al., 1989; Reid et al., 2011).

Even though trail shortening utilizes segments that were offered by individual ants, it is inherently different from the amplification schemes described in the previous section. This difference is manifested in the fact that in the previous examples require that the informed ant "know that she knows" and assess the quality of the information that she holds. Conversely, during trail shortening ants that mark short-cuts are not required to hold any knowledge about the quality of their solutions. Rather, it is the group that either amplifies or eventually ignores this alternative trail segments through a pheromone based positive-feedback mechanism.

## *Emergent collective cognition*

So far, we focused on group level behaviors that gain their computational power by amplifying the actions and decisions of individuals. This differs from the notion of emergent cognition wherein collective scale processes allow the group to *qualitatively* transcend individual capabilities.

An intuitive example for emergence is the different physical phases of matter. Here, minimal changes in temperature or the coupling between microscopic particles leads to qualitatively different macroscopic phases (*e.g.* the solid to liquid transition). It is therefore interesting to ask whether grouping together, not simple inanimate particles, but rather cognitive individuals with a memory and complex behavioral rules can be expected to display different and perhaps higher forms of emergence. Specifically, what forms of cognitive emergence occur in the case of social insects? In this section, we list several examples for emergent collective actions. As before, the examples are loosely ordered according to the increasing gaps between the individual and the group.

**Weighted response to multiple stimuli**
Since they are grouped into large ensembles individual insects are, inevitably, much smaller than the size of the colony and its territory. As a consequence, individuals cannot have direct access to large-scale environmental and internal colony conditions. Despite this, the colony as a whole must react to the full set of stimuli and appropriately divide the work force(Robinson, 1992).Models(Beshers and Fewell, 2001) suggestthat colony level division of labor can result from single insects with different task thresholds(Bonabeau, 1996)that resolve work demands which they locally experience(Franks and Tofts, 1994). Similar to the house-hunting example described above this colony level phenomenon relies on the cognitive assessments of individuals. The difference being that, in this case, the colony does not form a consensus around the solution of a single individual but rather divides the work force in a weighted manner according to information that is too spread out tobe available to any one individual. Division of labor can include more complex mechanisms, such as recruitment, that allow ants to employ not only personal but also social information in their decisions(Robinson et al., 2009a).

**Partial decoupling between individual and collective scales**
Cooperative transport is the process in which a group of ants retrieves a food item much too large for any of them to move on their own. During this process, the information available to individuals may be plainly misleading and counterproductive for the group's collective goals. This happens when trajectories to the nest as experienced by the small ants may be inaccessible to the large loads and even take it towards dead-ends that are difficult to escape. It was shown that to avoid such deadlocks, the macroscopic scale occasionally decouples from the possibly misleading information available at the microscopic scale(Fonio et al., 2016). This mechanism allows the group to utilize beneficial information while using noise present at the group level to escape deadlocks (local minima) and avoid the potentially devastating consequences of fully relying on misleading information. Importantly, such decoupling does not require that any single individual detect at any point in time, whether information is valuable or misleading.

Emergence in this system is evident as a separation between collective and individual behavior. This is evident, as the carrying group (and the food item they transport) does not follow a trajectory that was suggested by any single ant.

**Collective response independent of individual actions**
In some cases, the group can display effective reactions to stimuli that are not conceived by any individual. Such can be the case in which a cooperatively carrying group hits an obstacle. In such instances, instead of attempting to advance directly towards the nest, the group decouples from the actions of its individual members and goes into a perpendicular motion that takes the carried piece of food towards the edges of the obstacle(Gelblum et al., 2015). It was shown that this change in collective motion does not require that any individual ant be aware of the obstacle and individually change her behavior(Gelblum et al., 2016). Rather, the physical constraint induced by the obstacle directly affects the group as a whole. As a response, the group's mode of motion changes in a way that facilitates obstacle circumvention.

**Self-organization withouta blueprint**
A final example for emergent behavior involves nest construction. Social insects construct some of the most magnificent structures in the biological world(Theraulaz et al., 2003). This is done through a stigmergic process(Theraulaz and Bonabeau, 1999) in which individuals locally interact with features of the structure by adding (or removing, in the case of dug nests) building material to them(Franks and Deneubourg, 1997). This induces indirect communication as insects interact with the product of the action of their nest-mates. It was further shown that, in some cases, building materials are combined with a volatile construction pheromone(Khuong et al., 2016). This adds a temporal dimension to the physical structure and expands the possibilities for local rules and the complexity of the resulting structure(Khuong et al., 2016).

In contrast to trail formation, for example, where the result of the collective effort is a refinement of almost complete structures suggested by individual ants, nest construction creates structures which appear to be far from the capabilities of any individual. Indeed, in this high form of emergence, it is unlikely that individuals have a blueprint of the desired final product. Nevertheless, they follow local rules and such that their collective effort results in the construction of intricate nests.

## Discussion

The first three sections of this discussion follow the structure of the previous example section. We raise and then discuss some hypotheses regarding the prevalence of collective-scale behaviors which strongly rely on the cognitive capabilities of individuals.

### *Individual-based collective cognition*

An important factor to note is that many of the immediate requirements of the social group coincide with those of the individuals that comprise this group. Like the solitary insect, the colony must also scan the environment for food (Gordon, 1995), locate shelters (Franks et al., 2002), transport food (Gelblum et al., 2015), confront predators (Lamon and Topoff, 1981; Monographs, 2016), and care for brood (Siveter et al., 2014; Wang et al., 2015). As noted above, the cognitive abilities of individuals in a colony do not seem to greatly differ from those of their solitary counterparts. As such the cognition of individuals within a colony hold *direct* advantages to the group as a whole.The*Individual based collective cognition*section,above, providesexampleswherein the actions of a single individual suffice for directing the entire colony.

Generally speaking, amplifying of the behavior of an individual requires two colony level processes: First,the identification of the specific individual of interest and then the amplification of its behavior. For example, as mentioned above, the first ant to find a food source has the capacity to trace the complete path between the nest and the food(Cammaerts and Cammaerts, 1980; Hölldobler, 1976; Wilson, 1962). This ant makes itself identifiable by laying pheromone markings on the surface as it heads back to the nest(Beckers et al., 1992a). Amplification occurs as ants that follow this trail enforce it with further pheromonal markings. Finally, mass foraging develops along the trail drawn out by the initial recruiting ant(Beckers et al., 1989). In many cases, solutions provided by different individuals are in conflict such that one solution is amplified this must come at the expense of others. In this case a third decision making process occurs alongside identification and amplification. Some mechanisms by which such collective decisions occur include the preferential reinforcement of preferred solutions, such as preferred food sources being marked by higher pheromone concentrations (Beckers et al., 1992b; Jaffe and Howse, 1979; Sumpter and Beekman, 2003)and cross inhibition between alternative emerging solutions, as occurs during honeybee recruitment (Nieh, 2010).

### *Combining individual perspectives*

It is not always the case that an individual insect holds the complete solution to the colony's current needs. In fact, inherent constraints work to limit the value of individually held information. A first constraint involves the size of the individual when compared to that of its colony. Colonies reside over territories that are tens of meters (in the case of ants) or even kilometers (for wasps and bees) across and include large elaborate nest structures on a scale of several meters (Tschinkel, 2004). It is impossible for a single individual whose size is on the order of one centimeter to individually monitor these large areas. Other constraints are a single individual propensity to be badly informed, mistakenly wrong, or be limited by its cognitive capacity (Sasaki and Pratt, 2012).

The second set of examples as above highlights behaviors in which the group holds useful information about the environment but this information is distributed among a large number of individuals. Using communication, these information fragments can be integrated to yield collective decisions that take the "big picture" into account. Such integration can often be classified using two general schemes defined for animal groups in general: The "many eyes principle" (Ward et al., 2011) in which the group's capacity for surveillance increases with the number of alert animals and the "many wrongs principle"(Biro et al., 2006; Simons, 2004) in which averaging effects work to reduce "noise" at the scale of an individual animal to yield accurate collective action. Integration of information happens over a large number of contexts and species.Some recurring principles in this process include: distributed integration, nonlinearities (Sumpter and Beekman, 2003), positive and negative feedback loops (Franks et al., 2002; Nieh, 2010), use of interaction rates and cuing delays (Camazine, 1993; Greene et al., 2013), higher influence and lower propensity to be influenced by better informed individuals (Korman et al., 2014; Razin et al., 2013; Schultz et al., 2008), and correctly weighing individual vs. social information (Robinson et al., 2009b; Robinson et al., 2012).

This section also presents examples in which colonies have evolved to achieve a collective task by amplifying the behaviors of individuals that work towards the small scale version of this same task. For example, in the context of cooperative transport the actions of a leader ant are exactly those she would take when individually transporting a small food item to the nest. The group allows these actions to be the driving force behind the transport of a large heavy item.

### *Emergent group level cognition*

Last, some colony functions may fall outside of the solitary insect's behavioral or cognitive repertoire. These include large scale behaviors such as assessing the relation between the size of two objects when bothof these are much larger than the insect itself (Fonio et al., 2016; Gelblum et al., 2016), and active food dissemination (Camazine, 1993; Greenwald et al., 2015; Howard and Tschinek, 1981). In such cases, to achieve the group-level task, the relevant capabilitiesmust arise either by newly evolved individual traits or through a collective process that relies on the connectivity between individuals (a combination of these two options is also possible). Cases in which group level processes bestow the group with abilities that are qualitatively beyond those of its individual members can, by definition(De Wolf and Holvoet, 2005), be considered as a form of emergence.

In the two examples of cooperative transport presented above, group problem solving capabilities increase beyond those of individuals. In these cases, the small size of individual insect may prevent it from grasping the relevant large-scale relationships between the carried load and the obstacle. Therefore, in these cases the problem solving on the level of the group must decouple to some extent from the array of, possibly misleading, solutions as offered by individuals, even if these are highly adept navigators. The mechanisms that enable such decoupling between the scales do not have to be complex. In these examples, group-level noise in scent mark following (Fonio et al., 2016) and the persistence that results from alignment of forces (Gelblum et al., 2016) suffice for efficient transport that interacts with the environment on the relevant scale of the ant team and the large carried load.

Stigmergic nest construction holds the potential for higher levels of emergence. Here, nest construction is carried out by individuals who follow simple local rules while constantly reacting to the environmental product of their previous actions (Grassé, 1959). Following these stigmergic principles (Theraulaz et al., 2003) may allow insects to construct elaborate nest architectures (Tschinkel, 2004) without any blueprint. The degree of emergence in such processes is difficult to define. On the one hand, by using a stigmergic process, a single insect, a solitary queen for example, may construct a structure for which it holds no internal representation (Camazine et al., 2001; Theraulaz and Bonabeau, 1999). A larger group in which each individual follows the same rules may achieve faster construction but the quality of emergence remains constant (Theraulaz and Bonabeau, 1999). On the other hand, it has been shown that pheromone deposition onto the constructed nest is an essential part of collective stigmergic nest construction. The addition of this time dependent component (pheromones evaporate) has the potential of allowing the group to achieve more complex forms of emergence not achievable by a single individual working alone (Khuong et al., 2016). Indeed, such processes can be modeled using the mathematics of cellular automata a theory which holds the potential of describing extreme forms of emergence(Wolfram, 2002).

### *Prevalence of individual-based mechanisms*

Next, we raise somehypothesesin anattempt to provide rationale for the observed imminence of individuals to colony scale processes.

As shown above, it is often the case that colony level behaviors are a consequence of the direct of amplification of individual actionsthat rely on individual cognition. This didn't have to be the case. One could imagine an evolutionary pathway where collective cognition is constructed from a network of different individuals each manifesting different, possibly more basic, capabilities.In this case, the actions of individuals don't directly coincide with the actions of the group or even with each other. Rather, group performance emerges from the coordination between individuals. An example for this are neurons in the brain. Neurons have evolved to be cells whose actions are purely computational such that their spiking activity bears no direct relations to the collective process at which they participate. Hypothetically, there is no *a-priori* prevention that such structures arise in a social insect colony to serve as the basis for highly efficient collective scale behaviors. Why, therefore, is this not the general case?

We suggest that the first part of the answer to this question has to do with the availability and high quality performances that individual-based group level cognition can achieve. As stated above, by the time group cognition has evolved, individual insects were already developed independent organisms(Farris and Schulmeister, 2010; Ma et al., 2012). As such, they already possessed many of the cognitive resources that are required by the group. Furthermore, these individual capabilities are, in no sense, simple. For example, individual insects are highly adept navigators. To get from one place to another individual insects employ a toolbox consisting of multiple tactics, often applied in parallel. Across different insects, these include landmark navigation (Collett et al., 1993), dead reckoning (Collett and Collett, 2000), backtracking (Wystrach et al., 2013), and cognitive map (Gould, 1986)navigation by means of scents (Morgan, 2009), visual cues (Collett et al., 1993; Esch et al., 2001; Wehner, 2003), temperature, and even magnetic fields. Collective cognition that relies on such individual capabilities holds the advantage of utilizing these highly non-trivial traits.

The fact that individual insects possess high cognitive capabilities does not, however, suffice in explaining why these capabilities take a central role in many collective level processes. In the context of the navigation example as in the previous paragraph, it may very well be the case that a navigational toolbox that relies on the distributed actions of a large number of cooperating insects not only exists but, also, outperforms other, individual-based, schemes. Again, one may ask why this is not the general case and why the role of individuals has remained so pronounced throughout evolution. We hypothesize that the answer to this question has to do with the complexity of efficient distributed solutions given the inherent constraints that apply for a colony of autonomous individuals. We hypothesize that in many cases, simple distributed algorithms would not outperform what is readily achievable by amplifying individual cognition. We further hypothesize that distributed algorithms that do surpass simpler amplification processes may be expected to be highly complex and therefore difficult to evolve. Hence, the system may be trapped within local minima within the fitness landscape which utilize the individual cognitive components that have evolved at earlier stages. The next section discusses some of the non-trivial limitations on emergent behavioral solutions.

## *Computational constraints on emergent collective cognition*

Emergence is often associated with the notion of a group that "exceeds the sum of its parts". Theoretically speaking, achieving such highly effective cooperation typically requires a substantial degree of coordination. An intuitive example for this is the parallel search problem in which a group of non-communicating random walkers that start at a given location aim to collectively cover an area and then share the profits of their findings. In this context, to "exceed the sum of its parts" implies that the group of *N* searchers cover the area more than *N* times faster than each individual, were it acting on its own. In fact, in many natural topologies (Alon et al., 2011; Efremenko and Reingold, 2009), the situation is very far from this. For example, in grid topologies when searchers do not communicate and as long as N is not too small, multiple random walks that start from a single point, typically achieve negligible speed-up in cover time when compared to a single searcher(Alon et al., 2011) (*i.e., the group takes almost as much time to cover an area as a single one of its members*). In this case, being part of a group becomes highly unbeneficial from an individual's point of view: food must be shared but without the advantage of obtaining it at a faster rate.

Biological systems often achieve emergence using involved communication. The most celebrated example for this is that of the brain, viewed as a large ensemble of neurons. In the case of brains, such coordination heavily relies on the fact that the neurons are organized into stable networks such that the set of neighboring neurons of a given neuron remains, relatively, fixed. Synaptic plasticity allows a pair of neighboring neurons to undergo a mutual feedback process and fine-tune their connectivity(Hebb, 1949). This, as conceptually demonstrated by the Hopfield model (Hopfield, 1982), may provide the system with its immense computational power. Indeed, Hopfield networks have been shown to be capable of universal computation in the Turing sense (Síma and Orponen, 2003).

What can one expect for social insect colonies? On the one hand,these are very far from being non-communicating. On the other hand, interaction networks appear to be more loosely defined than those that characterize a brain. Indeed, in the insect colony fixed organization structures may be difficult to achieve for long periods of time, due to the inherent mobility and anonymity constraints. This lack of structure leads to a lack of knowledge that further constrains the system(Burgos and Polani, 2016). Namely, while learning in the brain happens by strengthening the synaptic connections between pairs of specific hard-wired (and therefore mutually identifiable) neurons, it is difficult imagine such mechanisms for freely moving, anonymous insects. Further, many brain functions are known to depend on precise interaction patterns that enable, for example, high levels of synchrony(Abeles, 2010). Such preciseness is difficult to imagine in the case of autonomous independent agents.

It appears that things become even worse if communication and information flows across the systems are, themselves, limited or distorted (Feinerman and Korman, 2012; Feinerman et al., 2014; Korman et al., 2014; Razin et al., 2013). The highly dynamic environment and interchanging network structures that characterize the social insect colony make it difficult to implement error correcting mechanisms such as repeatedly sending the same message in orderto reduce distortion. As demonstrated in (Feinerman and Korman, 2012; Feinerman et al., 2014; Korman et al., 2014; Razin et al., 2013) such circumstances make even basic distributed tasks, such as rumor spreading, challenging. It is therefore reasonable to assume that emergent phenomena would be even more difficult to implement in such conditions.

## Summary

Useful information exists at the level of the individual insect. This often suffices for the group's needs and indeed we have presented many examples in which the group follows individuals by amplifying their effect. In other cases, the inherent scale gap between individual and group may render information held by individuals partial, irrelevant, or even misleading to the collective goals of the group. In such cases, emergent collective phenomenon may kick in. Cognition that emerges at the level of the colony is subject to multiple constraints that mainly result from the fact that individuals maintain their autonomy as insects within the colony. Indeed, the most complex collective circuits described to date(Nieh, 2010; Pratt et al., 2005; Seeley et al., 2011) may be defined as simple when compared to the neural circuits that allow, for example, an ant to find her nest by using vector integration(Ofstad et al., 2011; Wehner, 2003).

Interestingly, the ways in which collective cognition appears, even during a single behavior, are not mutually exclusive.An example for thiscomes from cooperative transport in *Paratrechina longicornis* ants. Indeed, during this behavior the ants simultaneously benefit from both individual based(Gelblum et al., 2015) and collective based cognition(Fonio et al., 2016; Gelblum et al., 2016). This balance between the organizational scales allows the system to enjoy the best of both the macroscopic and the microscopic worlds.

Forming better connections between different collective behaviors and the levels of emergence that characterize them requires further research. We suggest that such research be focused on two avenues. The first is a computational study of the powers and limitations of natural distributed algorithms(Feinerman et al., 2014; Greenwald et al., 2015). Specifically, there is a need to better understand the complexity and evolvabilty of distributed solutions of different qualities. The second avenue involves empirical studies that attempt to trace the evolution of emergent, communication-based solutions. An example for this direction can be a comparative study of nest structures across a large number of phylogenetically related ant or termite species.

## Acknowledgements

This work has received funding from the European Research Council (ERC) under the European Union's Horizon 2020 research and innovation program (grant agreement No 648032).O.F., incumbent of the Louis and Ida Rich career development chair, supported by Israel Science Foundation grant 833/15, would like to thank and the Clore Foundation for their ongoing support. A.K. We would like to than Yossi Yovel for his comments on this manuscript.

# Bibliography

**Abeles, M.** (2010). Synfire Chains. In *Encyclopedia of Neuroscience*, pp. 829–832.

**Aiello, L. C. and Wheeler, P.** (1995). The Expensive-Tissue Hypothesis: The Brain and the Digestive System in Human and Primate Evolution. *Curr. Anthropol.* **36**, 199.

**Alloway, T.** (1972). Learning and memory in insects. *Annu. Rev. Entomol.* **17**, 43–56.

**Alon, N., Avin, C., Koucky, M., Kozma, G., Lotker, Z. and Tuttle, M. R.** (2011). Many Random Walks Are Faster Than One. *Comb. Probab. Comput.* **20**, 481–502.

**Amor, F., Ortega, P., Cerdá, X. and Boulay, R.** (2010). Cooperative prey-retrieving in the ant Cataglyphis floricola: an unusual short-distance recruitment. *Insectes Soc.* **57**, 91–94.

**Anderson, C. and McShea, D. W.** (2001). Individual versus social complexity, with particular reference to ant colonies. *Biol. Rev. Camb. Philos. Soc.* **76**, 211–37.

**Beckers, R., Goss, S., Deneubourg, J. L. and Pasteels, J.** (1989). Colony size, communication, and ant foraging strategy. *Psyche (Stuttg).* **96**, 239–256.

**Beckers, R., Deneubourg, J.-L. and Goss, S.** (1992a). Trail Laying Behavior During Food Recruitment In The Ant Lasius-Niger (L). *Insectes Soc.* **39**, 59–72.

**Beckers, R., Deneubourg, J. L. and Goss, S.** (1992b). Trails and U-turns in the selection of a path by the ant Lasius niger. *J. Theor. Biol.* **159**, 397–415.

**Beekman, M. and Dussutour, A.** (2007). How to Tell Your Mates Recruitment Mechanisms. 105–124.

**Behmer, S. T.** (2009). Animal Behaviour: Feeding the Superorganism. *Curr. Biol.* **19**, R366–R368.

**Beshers, S. N. and Fewell, J. H.** (2001). Models of division of labor in social insects. *Annu. Rev. Entomol.* **46**, 413–40.

**Biro, D., Sumpter, D. J. T., Meade, J. and Guilford, T.** (2006). From compromise to leadership in pigeon homing. *Curr. Biol.* **16**, 2123–8.

**Blackiston, D., Briscoe, A. D. and Weiss, M. R.** (2011). Color vision and learning in the monarch butterfly, Danaus plexippus (Nymphalidae). *J. Exp. Biol.* **214**, 509–520.

**Blonder, B. and Dornhaus, A.** (2011). Time-ordered networks reveal limitations to information flow in ant colonies. *PLoS One* **6**, e20298.

**Blum, M. S.** (1969). Alarm pheromones. *Annu. Rev. Entomol.* **14**, 57–80.

**Bonabeau, E.** (1996). Quantitative study of the fixed threshold model for the regulation of division of labour in insect societies. *Proc. R. Soc. B Biol. Sci.* **263**, 1565–1569.

**Brady, S. G., Schultz, T. R., Fisher, B. L. and Ward, P. S.** (2006). Evaluating alternative hypotheses for the early evolution and diversification of ants. *Proc. Natl. Acad. Sci.* **103**, 18172–18177.

**Brower, L.** (1996). Monarch butterfly orientation: missing pieces of a magnificent puzzle. *J. Exp. Biol.* **199**, 93–103.

**Burgos, A. C. and Polani, D.** (2016). An Informational Study of the Evolution of Codes and of Emerging Concepts in Populations of Agents. *Artif. Life* **22**, 196–210.

**Byrne, M., Dacke, M., Nordstrom, P., Scholtz, C. and Warrant, E.** (2003). Visual cues used by ball-rolling dung beetles for orientation. *J. Comp. Physiol. A Sensory, Neural, Behav. Physiol.* **189**, 411–418.

**Camazine, S.** (1993). The regulation of pollen foraging by honey bees: how foragers assess the colony's need for pollen. *Behav. Ecol. Sociobiol.* **32**, 265–272.

**Camazine, S., Deneubourg, J.-L., Franks, N. R., Sneyd, J., Theraulaz, G. and Bonabeau, E.** (2001). *Self-organisation in biological systems*.

**Cammaerts, M.-C. and Cammaerts, R.** (1980). Food recruitment strategies of the ants Myrmica sabuleti and Myrmica ruginodis. *Behav. Processes* **5**, 251–270.

**Chittka, L. and Niven, J.** (2009). Are bigger brains better? *Curr. Biol.* **19**, R995–R1008.

**Collett, M. and Collett, T. S.** (2000). How do insects use path integration for their navigation? *Biol. Cybern.* **83**, 245–59.

**Collett, T. S., Fry, S. N. and Wehner, R.** (1993). Sequence learning by honeybees. *J. Comp. Physiol. A* **172**, 693–706.

**Couzin, I. D.** (2009). Collective cognition in animal groups. *Trends Cogn. Sci.* **13**, 36–43.

**Crespi, B. J. and Yanega, D.** (1995). The definition of eusociality. *Behav. Ecol.* **6**, 109–115.

**Czaczkes, T. J. and Ratnieks, F. L. W.** (2013). Cooperative transport in ants ( Hymenoptera : Formicidae ) and elsewhere. *Myrmecological News* **18**, 1–11.

**De Wolf, T. and Holvoet, T.** (2005). Emergence versus self-organisation: Different concepts but promising when combined. *Lect. Notes Comput. Sci. (including Subser. Lect. Notes Artif. Intell. Lect. Notes Bioinformatics)* **3464 LNAI**, 1–15.

**Delattre, O., Sillam-Dussès, D., Jandák, V., Brothánek, M., Rücker, K., Bourguignon, T., Vytisková, B., Cvačka, J., Jiříček, O. and Šobotník, J.** (2015). Complex alarm strategy in the most basal termite species. *Behav. Ecol. Sociobiol.* 1945–1955.

**Deneubourg, J. L., Pasteels, J. M. and Verhaeghe, J. C.** (1983). Probabilistic behaviour in ants: A strategy of errors? *J. Theor. Biol.* **105**, 259–271.

**Dickson, B. J.** (2008). Wired for sex: the neurobiology of Drosophila mating decisions. *Science* **322**, 904–909.

**Dornhaus, A. and Franks, N. R.** (2008). Individual and collective cognition in ants and other insects (Hymenoptera: Formicidae). *Myrmecological News* **11**, 215–226.

**Edwards, S. C. and Pratt, S. C.** (2009). Rationality in collective decision-making by ant colonies. *Proc. Biol. Sci.* **276**, 3655–61.

**Efremenko, K. and Reingold, O.** (2009). How Well Do Random Walks Parallelize? In *Approximation, Randomization, and Combinatorial Optimization. Algorithms and Techniques*, pp. 476–489.

**Emerson, A. E.** (1939). Social Coordination and the Superorganism. *Am. Midl. Nat.* **21**, 182–209.

**Engel, M. S., Grimaldi, D. a. and Krishna, K.** (2009). Termites (Isoptera): Their Phylogeny, Classification, and Rise to Ecological Dominance. *Am. Museum Novit.* **3650**, 1–27.

**Esch, H. E., Zhang, S., Srinivasan, M. V and Tautz, J.** (2001). Honeybee dances communicate distances measured by optic flow. *Nature* **411**, 581–583.

**Evans, H. E.** (1966). The Behavior Patterns of Solitary Wasps. *Annu. Rev. Entomol.* **11**, 123–154.

**Farris, S. M. and Schulmeister, S.** (2010). Parasitoidism, not sociality, is associated with the evolution of elaborate mushroom bodies in the brains of hymenopteran insects. *Proc. Biol. Sci.* **278**, 940–51.

**Feinerman, O. and Korman, A.** (2012). Memory lower bounds for randomized collaborative search and implications for biology. In *Proceedings of International Symposium on Distributed COmputing (DISC)*, pp. 61–75.

**Feinerman, O. and Traniello, J. F. A.** (2015). Social complexity, diet, and brain evolution: modeling the effects of colony size, worker size, brain size, and foraging behavior on colony fitness in ants. *Behav. Ecol. Sociobiol.* **70**, 1063–1074.

**Feinerman, O., Haeupler, B. and Korman, A.** (2014). Breathe before speaking: efficient information dissemination despite noisy, limited, and anonymous communication. In *Proceedings of the 2014 ACM symposium on Principles of distributed computing - PODC '14*, pp. 114–123.

**Fewell, J. H.** (2003). Social insect networks. *Science* **301**, 1867–70.

**Feynmann, R. P.** (1985). The Amateur Scientist. In *Surely, you're joking Mr. Feynmann!*, p. 79. Bantam Books.

**Fonio, E., Heyman, Y., Boczkowski, L., Gelblum, A., Kosowski, A., Korman, A. and Feinerman, O.** (2016). A Locally-Blazed Ant Trail Achieves Efficient Collective Navigation Despite Limited Information. *Rev.*

**Franks, N. R.** (1989). Army Ants: A Collective Intelligence. *Am. Sci.* **77**, 138–145.

**Franks, N. and Deneubourg, J.** (1997). Self-organizing nest construction in ants: individual worker behaviour and the nest's dynamics. *Anim. Behav.* **54**, 779–96.

**Franks, N. R. and Tofts, C.** (1994). Foraging for work : how tasks allocate workers. *Anim. Behav.* **48**, 470–472.

**Franks, N. R., Pratt, S. C., Mallon, E. B., Britton, N. F. and Sumpter, D. J. T.** (2002). Information flow, opinion polling and collective intelligence in house-hunting social insects. *Philos. Trans. R. Soc. Lond. B. Biol. Sci.* **357**, 1567–83.

**Franks, N. R., Mallon, E. B., Bray, H. E., Hamilton, M. J. and Mischler, T. C.** (2003). Strategies for choosing between alternatives with different attributes: exemplified by house-hunting ants. *Anim. Behav.* **65**, 215–223.

**Gathmann, A. and Tscharntke, T.** (2002). Foraging ranges of solitary bees. *J. Anim. Ecol.* **71**, 757–764.

**Gelblum, A., Pinkoviezky, I., Fonio, E., Ghosh, A., Gov, N. and Feinerman, O.** (2015). Ant groups optimally amplify the effect of transiently informed individuals. *Nat. Commun.* **6**, 1–9.

**Gelblum, A., Pinkoviezky, I., Fonio, E., Gov, N. S. and Feinerman, O.** (2016). The Ant Pendulum: problem solving through an emergent oscillatory phase. *Proc. Natl. Acad. Sci.* **Accepted f**,.

**Gillooly, J. F., Hou, C. and Kaspari, M.** (2010). Eusocial insects as superorganisms Insights from metabolic theory. *Commun. Integr. Biol.* **3**, 360–362.

**Gordon, D. M.** (1995). The expandable network of ant exploration. *Anim. Behav.* **50**, 995–

1007.

**Gordon, D. M. and Mehdiabadi, N. J.** (1999). Encounter rate and task allocation in harvester ants. *Behav. Ecol. Sociobiol.* **45**, 370–377.

**Gordon, D. M., Paul, R. E. and Thorpe, K.** (1993). What is the function of encounter pattern in ant colonies. *Anim. Behav.* **45**, 1083–1100.

**Goss, S., Aron, S., Deneubourg, J. L. and Pasteels, J. M.** (1989). Self-organized shortcuts in the Argentine ant. *Naturwissenschaften* **76**, 579–581.

**Gould, J. L.** (1986). The locale map of honey bees: do insects have cognitive maps? *Science* **232**, 861–863.

**Grassé, P. P.** (1959). La reconstruction du nid et les coordinations interindividuelles chez Bellicositermes natalensis et Cubitermes sp. la th??orie de la stigmergie: Essai d'interpr??tation du comportement des termites constructeurs. *Insectes Soc.* **6**, 41–80.

**Greene, M. J., Pinter-Wollman, N. and Gordon, D. M.** (2013). Interactions with combined chemical cues inform harvester ant foragers' decisions to leave the nest in search of food. *PLoS One* **8**, e52219.

**Greenwald, E., Segre, E. and Feinerman, O.** (2015). Ant trophallactic networks: simultaneous measurement of interaction patterns and food dissemination. *Sci. Rep.* **5**, 12496.

**Hebb, D. O.** (1949). *The organization of behavior: a neuropsychological theory*.

**Holldobler, B.** (1971). Recruitment Behavior in Camponotus socius (Hym. Formicidae). *Z. Vgl. Physiol.* **75**, 123–142.

**Holldobler, B. and Wilson, E. O.** (1990). *The Ants*. Harvard University Press.

**Hölldobler, B.** (1976). Recruitment behavior, home range orientation and territoriality in harvester ants, Pogonomyrmex. *Behav. Ecol. Sociobiol.* **1**, 3–44.

**Hou, C., Kaspari, M., Vander Zanden, H. B. and Gillooly, J. F.** (2010). Energetic basis of colonial living in social insects. *Proc. Natl. Acad. Sci. U. S. A.* **107**, 3634–3638.

**Howard, R. W. and Blomquist, G. J.** (2005). Ecological, behavioral, and biochemical aspects of insect hydrocarbons. *Annu. Rev. Entomol.* **50**, 371–393.

**Howard, D. F. and Tschinek, W. R.** (1981). The flow of food in colonies of the fire ant, Solenopsis invicta: a multifactorial study. *Physiol. Entomol.* **6**, 297–306.

**Jaffe, K. and Howse, P. E.** (1979). The mass recruitment system of the leaf cutting ant, Atta cephalotes (L.). *Anim. Behav.* **27**, 930–939.

**Jandt, J. M. and Dornhaus, A.** (2009). Spatial organization and division of labour in the bumblebee Bombus impatiens. *Anim. Behav.* **77**, 641–651.

**Jeanson, R. and Deneubourg, J. L.** (2009). Positive feedback, convergent collective patterns, and social transitions in arthropods. In *Organization of Insect Societies - From genome to sociocomplexity*, pp. 460–482.

**Jones, J. C.** (2004). Honey Bee Nest Thermoregulation: Diversity Promotes Stability. *Science (80-. ).* **305**, 402–404.

**Khuong, A., Gautrais, J., Perna, A., Sbaï, C., Combe, M., Kuntz, P., Jost, C. and Theraulaz, G.** (2016). Stigmergic construction and topochemical information shape ant nest architecture. *Proc. Natl. Acad. Sci.* **113**, 1303–1308.

**King, H., Ocko, S. and Mahadevan, L.** (2015). Termite mounds harness diurnal temperature oscillations for ventilation. *Proc. Natl. Acad. Sci.* **112**, 11589–11593.

**Korman, A., Greenwald, E. and Feinerman, O.** (2014). Confidence Sharing: An Economic Strategy for Efficient Information Flows in Animal Groups. *PLoS Comput. Biol.* **10**, e1003862.

**Lamon, B. and Topoff, H.** (1981). Avoiding predation by army ants: Defensive behaviours of three ant species of the genus Camponotus. *Anim. Behav.* **29**, 1070–1081.

**Ma, X., Hou, X., Edgecombe, G. D. and Strausfeld, N. J.** (2012). Complex brain and optic lobes in an early Cambrian arthropod. *Nature* **490**, 258–261.

**Mailleux, A.-C., Deneubourg, J.-L. and Detrain, C.** (2003). Regulation of ants' foraging to resource productivity. *Proc. R. Soc. London. Ser. B Biol. Sci.* **270**, 1609–1616.

**Mallon, E. B. and Franks, N. R.** (2000). Ants estimate area using Buffon's needle. *Proc. Biol. Sci.* **267**, 765–70.

**Martin, S. and Drijfhout, F.** (2009). A review of ant cuticular hydrocarbons. *J. Chem. Ecol.* **35**, 1151–61.

**Mersch, D. P., Crespi, A. and Keller, L.** (2013). Tracking individuals shows spatial fidelity is a key regulator of ant social organization. *Science* **340**, 1090–3.

**Monographs, E.** (2016). Colony Defense Strategies of the Honeybees in Thailand Author ( s ): Thomas D . Seeley , Robin Hadlock Seeley and Pongthep Akratanakul Published by : Wiley Stable URL : http://www.jstor.org/stable/2937344 REFERENCES Linked references are available on JSTO. **52**, 43–63.

**Moreau, C. S.** (2006). Phylogeny of the Ants: Diversification in the Age of Angiosperms. *Science (80-. ).* **312**, 101–104.

**Moreau, M., Arrufat, P., Latil, G. and Jeanson, R.** (2011). Use of radio-tagging to map spatial organization and social interactions in insects. *J. Exp. Biol.* **214**, 17–21.

**Morgan, D. E.** (2009). Trail pheromones of ants. *Physiol. Entomol.* **34**, 1–17.

**Nieh, J. C.** (2010). A negative feedback signal that is triggered by peril curbs honey bee recruitment. *Curr. Biol.* **20**, 310–5.

**Noirot, C. and Darlington, J. P. E. C.** (2000). Termite Nests: Architecture, Regulation, and Defense. In *Termites: Evolution, Sociality, Symbiosis Ecology*, pp. 121–139. Springer.

**O'Neill, K.** (2001). *Solitary Wasps: Behavior and Natural History*.

**Ofstad, T. A., Zuker, C. S. and Reiser, M. B.** (2011). Visual place learning in Drosophila melanogaster. *Nature* **474**, 204–7.

**Pinter-Wollman, N., Wollman, R., Guetz, A., Holmes, S. and Gordon, D. M.** (2011). The effect of individual variation on the structure and function of interaction networks in harvester ants. *J. R. Soc. Interface* **8**, 1562–73.

**Pratt, S. C., Sumpter, D. J. T., Mallon, E. B. and Franks, N. R.** (2005). An agent-based model of collective nest choice by the ant Temnothorax albipennis. *Anim. Behav.* **70**, 1023–1036.

**Queller, D. C. and Strassmann, J. E.** (1998). Kin Selection and Social Insects. *Bioscience* **48**, 165–175.

**Ramsden, E., Adams, J., Holldobler, B., Shepherd, J. D., North, R. D., Jackson, C. W., Howse,**

**P. E., Bash, E., Millor, J., Am??, J. M., et al.** (2009). Orientation cues used by ants. *Insectes Soc.* **37**, 101–115.

**Raw, A.** (1972). The biology of the solitary bee Osmia rufa (L.) (Megachilidae). *Trans. R. Entomol. Soc. London* **124**, 213–229.

**Razin, N., Eckmann, J. and Feinerman, O.** (2013). Desert ants achieve reliable recruitment across noisy interactions. *J. R. Soc. Interface* **10**, 20130079.

**Reid, C. R., Sumpter, D. J. T. and Beekman, M.** (2011). Optimisation in a natural system: Argentine ants solve the Towers of Hanoi. *J. Exp. Biol.* **214**, 50–8.

**Richardson, T. O. and Gorochowski, T. E.** (2015). Beyond contact-based transmission networks : the role of spatial coincidence. *J. R. Soc. Interface* **12**, 20150705.

**Richardson, T. O., Christensen, K., Franks, N. R., Jensen, H. J. and Sendova-Franks, A. B.** (2011). Ants in a labyrinth: A statistical mechanics approach to the division of labour. *PLoS One* **6**,.

**Rieucau, G. and Giraldeau, L.-A.** (2011). Exploring the costs and benefits of social information use: an appraisal of current experimental evidence. *Philos. Trans. R. Soc. Lond. B. Biol. Sci.* **366**, 949–57.

**Robinson, G. E.** (1992). Regulation of division of labor in insect societies. *Annu. Rev. Entomol.* **37**, 637–665.

**Robinson, E. J. H., Feinerman, O. and Franks, N. R.** (2009a). Flexible task allocation and the organization of work in ants. *Proc. Biol. Sci.* **276**, 4373–80.

**Robinson, E. J. H., Richardson, T. O., Sendova-Franks, A. B., Feinerman, O. and Franks, N. R.** (2009b). Radio tagging reveals the roles of corpulence, experience and social information in ant decision making. *Behav. Ecol. Sociobiol.* **63**, 627–636.

**Robinson, E. J. H., Franks, N. R., Ellis, S., Okuda, S. and Marshall, J. a R.** (2011). A simple threshold rule is sufficient to explain sophisticated collective decision-making. *PLoS One* **6**, e19981.

**Robinson, E. J. H., Feinerman, O. and Franks, N. R.** (2012). Experience, corpulence and decision making in ant foraging. *J. Exp. Biol.* **215**, 2653–9.

**Robinson, E. J. H., Feinerman, O. and Franks, N. R.** (2014). How collective comparisons emerge without individual comparisons of the options. *Proc. Biol. Sci.* **281**,.

**Robson, S. K. and Traniello, J. F.** (2002). Transient division of labor and behavioral specialization in the ant Formica schaufussi. *Naturwissenschaften* **89**, 128–131.

**Roces, F., Tautz, J. and Hölldobler, B.** (1993). Stridulation in leaf-cutting ants - Short-range recruitment through plant-borne vibrations. *Naturwissenschaften* **80**, 521–524.

**Sasaki, T. and Pratt, S. C.** (2011). Emergence of group rationality from irrational individuals. *Behav. Ecol.* **22**, 276–281.

**Sasaki, T. and Pratt, S. C.** (2012). Groups have a larger cognitive capacity than individuals. *Curr. Biol.* **22**,.

**Sasaki, T., Granovskiy, B., Mann, R. P., Sumpter, D. J. T. and Pratt, S. C.** (2013). Ant colonies outperform individuals when a sensory discrimination task is difficult but not when it is easy. *Proc. Natl. Acad. Sci.* **110**, 13769–13773.

**Schmidt, J. O.** (1990). *Insect defenses: adaptive mechanisms and strategies of prey and*

*predators.* SUNY Press.

**Schultz, K. M., Passino, K. M. and Seeley, T. D.** (2008). The mechanism of flight guidance in honeybee swarms: subtle guides or streaker bees? *J. Exp. Biol.* **211**, 3287–95.

**Seeley, T. D.** (1996). *The Wisdom of the Hive*.

**Seeley, T. D., Visscher, P. K. and Passino, K. M.** (2006). Group Decision Making in Honey Bee Swarms. *Am. Sci.* **94**, 220–229.

**Seeley, T. D., Visscher, P. K., Schlegel, T., Hogan, P. M., Franks, N. R. and Marshall, J. A. R.** (2011). Stop Signals Provide Cross Inhibition in Collective Decision-Making by Honeybee Swarms. *Science*.

**Sendova-Franks, a. B. and Franks, N. R.** (1995). Spatial relationships within nests of the antLeptothorax unifasciatus(Latr.) and their implications for the division of labour. *Anim. Behav.* **50**, 121–136.

**Shultz, S. and Dunbar, R.** (2010). Encephalization is not a universal macroevolutionary phenomenon in mammals but is associated with sociality. *Proc. Natl. Acad. Sci. U. S. A.* **107**, 21582–21586.

**Simon, T. and Hefetz, A.** (1992). Dynamics of mass recruitment and its regulation in Tapinoma simrothi. *Biol. Evol. Soc. Insects* 325–334.

**Simons, A. M.** (2004). Many wrongs: the advantage of group navigation. *Trends Ecol. Evol.* **19**, 453–5.

**Siveter, D. J., Tanaka, G., Farrell, U. C., Martin, M. J., Siveter, D. J. and Briggs, D. E. G.** (2014). Exceptionally preserved 450-million-year-old ordovician ostracods with brood care. *Curr. Biol.* **24**, 801–806.

**Starks, P. T., Blackie, C. A. and Thomas D Seeley, P. T.** (2000). Fever in honeybee colonies. *Naturwissenschaften* **87**, 229–231.

**Strausfeld, N. J.** (1976). *Atlas of an insect brain.* Springer Science & Business Media.

**Stroeymeyt, N., Franks, N. R. and Giurfa, M.** (2011). Knowledgeable individuals lead collective decisions in ants. *J. Exp. Biol.* **214**, 3046–3054.

**Sumpter, D. J. T.** (2006). The principles of collective animal behaviour. *Philos. Trans. R. Soc. Lond. B. Biol. Sci.* **361**, 5–22.

**Sumpter, D. J. T. and Beekman, M.** (2003). From nonlinearity to optimality: pheromone trail foraging by ants. *Anim. Behav.* **66**, 273–280.

**Theraulaz, G. and Bonabeau, E.** (1999). A Brief History of Stigmergy. *Artif. Life* **5**, 97–116.

**Theraulaz, G., Gautrais, J., Camazine, S. and Deneubourg, J.-L.** (2003). The formation of spatial patterns in social insects: from simple behaviours to complex structures. *Philos. Trans. A. Math. Phys. Eng. Sci.* **361**, 1263–82.

**Tibbetts, E. a** (2002). Visual signals of individual identity in the wasp Polistes fuscatus. *Proc. Biol. Sci.* **269**, 1423–1428.

**Tschinkel, W. R.** (2004). The nest architecture of the Florida harvester ant, Pogonomyrmex badius. *J. Insect Sci.* **4**, 21.

**van Wilgenburg, E., Symonds, M. R. E. and Elgar, M. A.** (2011). Evolution of cuticular hydrocarbon diversity in ants. *J. Evol. Biol.* **24**, 1188–1198.

**Visscher, P. K.** (2007). Group decision making in nest-site selection among social insects. *Annu. Rev. Entomol.* **52**, 255–75.

**Von Frisc, K.** (1950). *Bees: their vision, chemical senses, and language*. Cornell University Press.

**Wang, B., Xia, F., Wappler, T., Simon, E., Zhang, H., Jarzembowski, E. A. and Szwedo, J.** (2015). Brood care in a 100-million-year-old scale insect. *Elife* **2015**, 4–11.

**Ward, A. J. W., Herbert-read, J. E., Sumpter, D. J. T. and Krause, J.** (2011). Fast and accurate decisions through collective vigilance in fish shoals. *Proc. Natl. Acad. Sci.* **108**, E27–E27.

**Waters, J. S., Holbrook, C. T., Fewell, J. H. and Harrison, J. F.** (2010). Allometric scaling of metabolism, growth, and activity in whole colonies of the seed-harvester ant Pogonomyrmex californicus. *Am. Nat.* **176**, 501–10.

**Wehner, R.** (2003). Desert ant navigation: how miniature brains solve complex tasks. *J. Comp. Physiol. A. Neuroethol. Sens. Neural. Behav. Physiol.* **189**, 579–88.

**Wilson, E. O.** (1962). Chemical communication among workers of the fire ant Solenopsis saevissima (Fr. Smith) 1. The Organization of Mass-Foraging. *Anim. Behav.* **10**, 134–147.

**Wilson, E. O. and Hölldobler, B.** (1988). Dense heterarchies and mass communication as the basis of organization in ant colonies. *Trends Ecol. Evol. (Personal Ed.* **3**, 65–8.

**Wilson, E. and Hölldobler, B.** (2009). *The superorganism: The Beauty, Elegance, and Strangeness of Insect Societies*. W. W. Norton & Company.

**Witte, V., Attygalle, A. B. and Meinwald, J.** (2007). Complex chemical communication in the crazy ant Paratrechina longicornis Latreille (Hymenoptera: Formicidae). *Chemoecology* **17**, 57–62.

**Wolfram, S.** (2002). *A New Kind Of Science*. Wolfram Media.

**Wystrach, A., Schwarz, S., Baniel, A. and Cheng, K.** (2013). Backtracking behaviour in lost ants: an additional strategy in their navigational toolkit. *Proc. Biol. Sci.* **280**, 20131677.